\documentclass[aps,pra,twocolumn,showpacs,superscriptaddress,footinbib,letterpaper,11pt,reprint]{revtex4-1}

\usepackage{graphicx}
\usepackage{epsfig}
\usepackage{epstopdf}

\usepackage{amsmath}
\usepackage{amsfonts}
\usepackage{amssymb}
\usepackage{bm}

\newcommand{\0}{\textsc{\tiny{(0)}}}

\newcommand{\bra}[1]{\ensuremath{\left\langle#1\right|}}
\newcommand{\ket}[1]{\ensuremath{\left|#1\right\rangle}}

\def\gsim{\lower.35em\hbox{$\stackrel{\textstyle>}{\textstyle\sim}$}}
\def\lsim{\lower.35em\hbox{$\stackrel{\textstyle<}{\textstyle\sim}$}}

\usepackage[utf8]{inputenc}
\usepackage[T1]{fontenc}

\usepackage{appendix}
\usepackage{color}
\definecolor{blue(pigment)}{rgb}{0.1, 0.3, 0.6}
\usepackage{url}
\usepackage[hyperfootnotes=false,hypertexnames=false]{hyperref}
\hypersetup
{
	pdftitle={Electro-Optics of Current-carrying Graphene},
	bookmarksnumbered=true,
	bookmarksopen=true,
	bookmarksopenlevel=1,
	colorlinks=true,
	citecolor=blue(pigment),
	linkcolor=blue(pigment),
	urlcolor=blue(pigment)
}

\begin{document}

\title{Electro-Optics of Current-carrying Graphene}

\author{Mohsen Sabbaghi}
\affiliation{Department of Electrical Engineering, University of Wisconsin-Milwaukee, Milwaukee, WI 53211, USA}

\author{Hyun-Woo Lee}
\email{hwl@postech.ac.kr}
\affiliation{Department of Physics, Pohang University of Science and Technology, Pohang 37673, Republic of Korea}

\author{Tobias Stauber}
\email{tobias.stauber@csic.es}
\affiliation{Materials Science Factory, Instituto de Ciencia de Materiales de Madrid, CSIC, E-28049 Madrid, Spain}

\date{\today}

\begin{abstract}
Electro-optical response of a current-carrying monolayer graphene is studied theoretically. Our calculation takes into account full (diagonal and non-diagonal) conductivity tensor obtained from a particle-conserving out-of-equilibrium distribution function of doped graphene. Our analytical and numerical results indicate that the presence of a moderate DC current throughout a doped graphene channel induces large Kerr rotations within a frequency range which can be tuned up to the mid-infrared frequency range.
\end{abstract}



\maketitle


\section{Introduction}\label{section:INTRODUCTION}
After a decade-long ubiquity of graphene, the electromagnetic (EM) response of this one-atom-thick honeycomb crystal of carbon atoms, in its current-carrying state, has only recently become the focus of increasing attention \cite{PhysRevB.81.115413,CHOI,PhysRevB.89.195447,Cheng:14,Behnaam15,2015arXiv151209044B,Duppen16,PhysRevB.94.035439,PhysRevLett.119.133901,2040-8986-19-6-065505,PhysRevB.97.085419,PhysRevB.97.165424}. Besides lattice effects \cite{PhysRevB.82.155412,PhysRevLett.106.045504} that are negligible in the optical limit, graphene is expected to exhibit an isotropic EM response. The presence of an in-plane field of uniaxial strain \cite{Pellegrino10,Pellegrino11,Martinez12} or a perpendicular magnetic field \cite{Crassee,Fialkovsky2012,Shimano,Zhou,Ellis,photonics2010013,PhysRevB.92.125426,Poumirol17}, however, breaks this isotropy and turns graphene into a birefringent optical medium.

Faraday rotations up to $6^{\circ}$ have been achieved upon transmission of linearly-polarized THz radiation through graphene under a perpendicularly-applied magnetic field of $\mathrm{B}_{\! \perp}\!=\!7\hspace{0.08em}\mathrm{T}$ at temperatures of $5\hspace{0.08em}\mathrm{K}$ \cite{Crassee} and $250\hspace{0.08em}\mathrm{K}$ \cite{Poumirol17}$\phantom{}$. Such large magneto-optical rotations, however, mainly occur at frequencies lying within the far-infrared (THz) band of EM spectrum, i.e., $1\hspace{0.08em}\mathrm{meV} \! < \! \hbar\omega \! < \! 80\hspace{0.08em}\mathrm{meV}$ \cite{Crassee}. In addition, magneto-optical phenomena are not the most suitable tool to achieve optical non-reciprocity (ONR) in integrated Photonics mainly because the undesirable impact of the magnetic field on the functionality of the nearby optical or electronic components cannot be avoided in sub-micron scales \cite{Lipson,Kamal,Shiyue,2015arXiv151209044B}.

The perpendicular static magnetic field also breaks the time reversal symmetry (TRS) of the nonlocal EM response of graphene leading to the emergence of edge magneto-plasmons \cite{Fet85,Wan12,Jin17}. Additionally, it has recently been shown that valley-selective population inversion in gapped Dirac materials (GDMs) such as biased bilayer graphene or transition metal dichalcogenides (TMDs) \cite{PhysRevLett.105.136805,Mak1489,SongKats} under optical pumping of circularly-polarized light yields a nonvanishing Berry flux, thus leading to broken TRS and the emergence of chiral (nonreciprocal) Berry plasmons (CBPs) \cite{Kum16,Son16}.

The presence of DC current has been predicted to cause the EM response of graphene to lose its invariance under (i) rotation in the local (optical) limit \cite{PhysRevB.81.115413,PhysRevB.89.195447,Behnaam15,2015arXiv151209044B,Duppen16,PhysRevB.94.035439} and (ii) time reversal (TR) in the nonlocal limit leading to different plasmonic group velocities depending on the direction of the external DC current \cite{PhysRevB.89.195447,Behnaam15,Duppen16,2015arXiv151209044B,PhysRevB.97.085419}. In this work, we show (i) how the presence of DC electric current in doped graphene breaks the rotational symmetry, (ii) how the resulting anisotropy leads to the emergence of off-diagonal elements of the conductivity tensor, and (iii) how such off-diagonal elements bring about electro-optical phenomena such as Kerr/Faraday rotation within a frequency range which can be tuned up to mid-infrared via the application of the gate and drain-source voltages.

This paper is structured as follows: Sec. \ref{section:EM_RESPONSE_OUT_OF_EQUILIBRIUM} provides details on the computation of the conductivity tensor of a driven $\pi$ electron gas and introduces the model to describe the nonequilibrium (NE) occupation of a driven electron gas. In Sec. \ref{section:THE_OPTICAL_RESPONSE}, we present analytic expressions for the optical conductivity of current-carrying graphene and discuss the scattering of light off current-carrying graphene. A summary, along with some concluding remarks, is given in Sec. \ref{section:SC}. Additional details are delivered through five appendices.

\section{EM response out of equilibrium}\label{section:EM_RESPONSE_OUT_OF_EQUILIBRIUM}
In this section, we will outline the basic steps of our theory by first defining the full response out of equilibrium. We will then introduce the shifted Fermi disk model which we shall use throughout this work.
\subsection{The conductivity tensor of Dirac fermions}\label{subsection:CONDUCTIVITY_TENSOR}
In response to an EM  perturbation, of frequency $\omega$ and in-plane wavevector $\mathbf{q} = \mathrm{q}_{x} \hat{\mathbf{e}}_{x} + \mathrm{q}_{y} \hat{\mathbf{e}}_{y}$, given by $\mathbf{\mathbf{E}}(\mathbf{r},t)\!=\!\overline{\mathbf{\mathbf{E}}}(\mathbf{q},z,\omega) \, e^{i[\mathrm{q}_{x} x + \mathrm{q}_{y} y- \omega t]}$, the $\pi$ electron gas in graphene undergoes current density oscillations given by $\mathbf{\mathbf{J}}(\mathbf{r},t)\!=\!\overline{\mathbf{\mathbf{J}}}(\mathbf{q},\omega) \delta_{\tiny{\text{D}}}\!\left(z\right) e^{i[\mathrm{q}_{x} x + \mathrm{q}_{y} y - \omega t]}$, with $\delta_{\tiny{\text{D}}}$ being the Dirac delta function. Such EM response can be described through the surface conductivity tensor $\tensor{\sigma}(\mathbf{q},\omega)$ which is defined via $\overline{\mathbf{\mathbf{J}}}(\mathbf{q},\omega) \! = \! \tensor{\sigma}(\mathbf{q},\omega) \cdot \overline{\mathbf{\mathbf{E}}}(\mathbf{q},z\!=\!0,\omega)$. For an isotropic sample, the optical conductivity tensor has a scalar nature, i.e., $\tensor{\sigma}^{\0}(\mathrm{q}\!=\!0,\omega) \! = \! \sigma^{\0}\!(\omega) \tensor{\mathrm{I}}$, where $\tensor{\mathrm{I}} \!\! \equiv \! \hat{\mathbf{\mathbf{e}}}_{x}\hat{\mathbf{\mathbf{e}}}_{x} \! + \hat{\mathbf{\mathbf{e}}}_{y}\hat{\mathbf{\mathbf{e}}}_{y}$ denotes the dyadic unit tensor, and $\sigma^{\0}\!(\omega)$ denotes the equilibrium-state optical conductivity that has been extensively studied in the literature \cite{1367-2630-8-12-318,Falkovsky2007,PhysRevB.76.153410,PhysRevLett.100.117401,PhysRevLett.101.196405,Nair1308, 1742-6596-129-1-012004,PhysRevB.78.085432,PhysRevB.83.165113,PhysRevB.93.125413,Stauber15}. In the general case where no isotropy is assumed, the components of the optical conductivity tensor of $\pi$ electron gas $\sigma_{n,\overline{n}}(\mathbf{q},\omega) \! \equiv \! \hat{\mathbf{\mathbf{e}}}_{n} \! \cdot  \tensor{\sigma}(\mathbf{q},\omega) \! \cdot \hat{\mathbf{\mathbf{e}}}_{\bar{n}}$ (where $n,\overline{n} \! = \! x,y$) should be obtained from the following summation over the first Brillouin zone (FBZ) \cite{Falkovsky2007,PhysRevB.93.125413}:
\begin{equation}\label{CONDUCTIVITY_TENSOR}
\sigma_{n,\overline{n}}(\mathbf{q},\omega) = 4 i g_{s} \gamma^{2} \sigma_{\!\text{u}} \sum_{\mathbf{k} \in \mathrm{FBZ}}{\mathrm{L}^{\mathbf{k},\mathbf{q}}_{n,\overline{n}}\left(\omega + i0^{\scriptscriptstyle{+}}\right)},
\end{equation}
where $\sigma_{\!\text{u}} \! \equiv \! \frac{e^{2}}{4\hbar}$ is the unit in terms of which the conductivity data in this work will be presented, $e$ denotes the elementary electric charge, $g_{s}\!=\!2$ is the spin degeneracy, $\gamma$ is the slope of Dirac cones and
\begin{equation}\label{CONDUCTIVITY_TENSOR_SUMMAND}
\mathrm{L}^{\mathbf{k},\mathbf{q}}_{n,\overline{n}}\left(\omega\right) \equiv \sum_{s,\overline{s}=\pm}{ \frac{n_{\text{\tiny{F}}}[\mathrm{E}^{\overline{s}}_{\mathbf{k}+\mathbf{q}}] - n_{\text{\tiny{F}}}[\mathrm{E}^{s}_{\mathbf{k}}]}{\mathrm{E}^{\overline{s}}_{\mathbf{k}+\mathbf{q}} - \mathrm{E}^{s}_{\mathbf{k}} -\hbar\omega}\frac{\mathrm{F}_{\!n,\overline{n}}^{s\overline{s}}\! \left(\mathbf{k},\mathbf{q}\right)}{\mathrm{E}^{\overline{s}}_{\mathbf{k}+\mathbf{q}} - \mathrm{E}^{s}_{\mathbf{k}}}},
\end{equation}
with $n_{\text{\tiny{F}}}[E]$ denoting the Fermi-Dirac (FD) distribution function
\begin{equation}\label{Fermi-Dirac}
n_{\text{\tiny{F}}}[E]=\left[1+\exp{\!\left(\frac{E-E_{\text{\tiny{F}}}}{k_{\text{\tiny{B}}}T_{e}}\right)}\right]^{-1},
\end{equation}
where $k_{\text{\tiny{B}}}$, $T_{e}$ and $E_{\text{\tiny{F}}}$ respectively denote the Boltzmann constant, the temperature and the Fermi energy of the $\pi$ electron gas. The function $\mathrm{E}^{s}_{\mathbf{k}}$ yields the energy eigen-value of the $\ket{\mathbf{k},s}$ eigen-state of the conduction ($s\!=\!+1$) or valence ($s\!=\!-1$) band. In addition, $\mathrm{F}_{\!n,\overline{n}}^{s\overline{s}}\!\left(\mathbf{k},\mathbf{q}\right)$ denotes the band overlap integral corresponding to the intraband ($s\overline{s}\!=\!+1$) or interband ($s\overline{s}\!=\!-1$) transitions.

The current-induced modification to the conductivity tensor $\Delta \tensor{\sigma} \equiv \tensor{\sigma}-\tensor{\sigma}^{\0}$ is the quantity of interest here and is solely determined by the eigen-states within a narrow neighborhood of the Fermi energy. To focus on these ``near-Fermi-level'' eigen-states, the summation over FBZ which yields $\Delta \tensor{\sigma}$ should be reduced into a polar integral around the Dirac point $\mathbf{k}_{D}$ via redefining the crystal momentum $\mathbf{k}$ to $\mathbf{k}+\mathbf{k}_{D}$ where $\mathbf{k}=\mathrm{k} \left[ \hat{\mathbf{\mathbf{e}}}_{x} \cos{\theta_{\mathbf{k}}} + \hat{\mathbf{\mathbf{e}}}_{y} \sin{\theta_{\mathbf{k}}} \right]$.

The application of tight-binding (TB) model within the Dirac cone approximation \cite{PhysRevB.81.085409} yields $\mathrm{E}^{s}_{\mathbf{k}} \!\cong\! s \gamma \mathrm{k}$ ($\gamma \! \equiv \! \frac{3at}{2}$) with $t \! \approx \! 2.7\hspace{0.08em}\mathrm{eV}$ and $a \! \approx \! 0.142\hspace{0.08em}\mathrm{nm}$ being the nearest-neighbor hopping amplitude and carbon-carbon bond length. Without the negligible lattice effects \cite{PhysRevB.82.155412}, the TB model yields
\begin{equation}\label{BAND_OVERLAP_INTEGRAL_TB}
\mathrm{F}_{\!n,\overline{n}}^{s\overline{s}}\!\left(\mathbf{k},\mathbf{q}\right) \! \cong \! \bra{s,\mathbf{k}} \tau_{n} \! \ket{\mathbf{k}\!+\!\mathbf{q},\overline{s}} \bra{\overline{s},\mathbf{k}\!+\!\mathbf{q}} \tau_{\overline{n}} \! \ket{\mathbf{k},s},
\end{equation}
where $\tau_{n}$ ($n\!=\!x,y$) denotes the $2 \! \times \! 2$ Pauli matrices \cite{1367-2630-8-12-318,VIGNALE,PhysRevB.80.075418}. Within the Dirac cone approximation, $\mathrm{F}_{\!n,\overline{n}}^{s\overline{s}}\! \left(\mathbf{k},\mathbf{q}\right)$ is specifically given by \cite{PhysRevB.80.075418,PhysRevB.82.155412}:
\begin{eqnarray}
&\displaystyle{2 \, \mathrm{F}_{\!x,x}^{s\overline{s}} \! \left(\mathbf{k},\mathbf{q}\right) \cong 1 + s\overline{s} \, \cos{[\theta_{\mathbf{k}}+\theta_{\mathbf{k}+\mathbf{q}}]}\,}\label{BAND_OVERLAP_INTEGRAL_XX}
\\[0.5ex]
&\displaystyle{2 \, \mathrm{F}_{\!x,y}^{s\overline{s}} \! \left(\mathbf{k},\mathbf{q}\right) = 2 \mathrm{F}_{\!y,x}^{s\overline{s}} \! \left(\mathbf{k},\mathbf{q}\right) \cong s\overline{s} \, \sin{[\theta_{\mathbf{k}}+\theta_{\mathbf{k}+\mathbf{q}}]}}\label{BAND_OVERLAP_INTEGRAL_XY}
\\[0.5ex]
&\displaystyle{2 \, \mathrm{F}_{\!y,y}^{s\overline{s}} \! \left(\mathbf{k},\mathbf{q}\right) \cong 1 - s\overline{s} \, \cos{[\theta_{\mathbf{k}}+\theta_{\mathbf{k}+\mathbf{q}}]}} . \label{BAND_OVERLAP_INTEGRAL_YY}
\end{eqnarray}
In the absence of DC electric current, $\tensor{\sigma}^{\0}(\omega)$ can be computed via plugging the FD distribution function with $E_{\text{\tiny{F}}}^{\0} \!=\! \pm \gamma  \sqrt{\pi n_{s}}$ as its Fermi energy, denoted by $n_{\text{\tiny{F}}}^{\0}[E]$, into Eq. (\ref{CONDUCTIVITY_TENSOR}), with $n_{s}$ denoting the density of injected ($E^{\0}_{\text{\tiny{F}}}\!\!>\!0$) or depleted ($E^{\0}_{\text{\tiny{F}}}\!\!<\!0$) electrons. The application of drain-source voltage along graphene channel pushes the $\pi$ electron gas out of its equilibrium. As in Refs. \citenum{PhysRevB.81.115413,Behnaam15} and \citenum{Duppen16}, in this work the nonequilibrium (NE) conductivity is obtained via feeding the NE distribution of the driven $\pi$ electron gas into the expression given by Eq. (\ref{CONDUCTIVITY_TENSOR_SUMMAND}).
\subsection{The shifted Fermi disk (SFD) model}\label{subsection:SFD}
The NE distribution of current-carrying $\pi$ electron gas can, in principle, be obtained via solving the Boltzmann transport equation (BTE) \cite{Levinson,MAHAN_BTE,Vasko}. We instead employ the phenomenological shifted Fermi disk (SFD) model \cite{Meric,DONHEE,Hosang,Behnaam15} which simulates the DC flux with a shift of the Fermi disk $\mathbf{k}_{\text{shift}}$ with respect to the Dirac point which can be related to the external DC electric field $\mathbf{E}_{\text{\tiny{DC}}}$ via $\mathbf{k}_{\text{shift}}=-\frac{e}{\hbar} \tau_{\text{\tiny{DC}}}\mathbf{E}_{\text{\tiny{DC}}}$ with $\tau_{\text{\tiny{DC}}}$ being the $\mathrm{DC}$ relaxation time. The SFD model is a particle-conserving model, meaning that the size of the Fermi disk and therefore, the electron density, $n_{s}$, are not affected by $\mathbf{E}_{\text{\tiny{DC}}}$. This model formulates the NE Fermi energy as follows \cite{Behnaam15}:
\begin{equation}\label{Nonequilibrium_EF}
E_{\text{\tiny{F}}}(\theta_{\mathbf{k}},\theta_{d})=E_{\text{\tiny{F}}}^{\0} \left[ \beta_{d} \cos{\theta} + \left[1\!-\!\beta^{2}_{d} \sin^{2}{\!\theta}\right]^{\!\frac{1}{2}} \right],
\end{equation}
where $\beta_{d} \equiv \mathrm{k}_{\text{shift}}/k_{\text{\tiny{F}}}^{\0} \leq 1$ is a dimensionless parameter expressing the shift of the Fermi disk $\mathrm{k}_{\text{shift}}$ in reciprocal space in units of the Fermi wave vector $k_{\text{\tiny{F}}}^{\0} \! \equiv \! |E_{\text{\tiny{F}}}^{\0}|/\gamma$, and $\theta \equiv \theta_{\mathbf{k}}\!-\!\theta_{d}$ with $\theta_{d}$ being defined through
\begin{equation}\label{DIMENSIONLESS_SHIFT_PARAMETER}
\boldsymbol{\beta}_{\!d} \equiv \frac{\mathbf{k}_{\text{shift}}}{k_{\text{\tiny{F}}}^{\0}} \equiv \beta_{d} \left[ \hat{\mathbf{\mathbf{e}}}_{x} \cos{\theta_{d}} + \hat{\mathbf{\mathbf{e}}}_{y} \sin{\theta_{d}} \right].
\end{equation}
The NE electronic occupation is then approximated by the FD distribution function $n_{\text{\tiny{F}}}[E]$ fed with an NE Fermi energy of $E_{\text{\tiny{F}}}(\theta_{\mathbf{k}},\theta_{d})$. Thus, the drift velocity, $\mathbf{v}_{\!d}\!=\!\left( \mathrm{v}_{\! d} / \mathrm{k}_{\text{shift}}\right) \, \mathbf{k}_{\text{shift}}$, reads as (see Appendix \ref{Appendix:A}):
\begin{equation}\label{Drift_Velocity}
\frac{\mathbf{v}_{\!d}}{v_{\text{\tiny{F}}}} = \frac{2\,\text{sgn}[E_{\text{\tiny{F}}}^{\0}]}{\pi} \boldsymbol{\beta}_{\!d} \int_{0}^{\pi} \left[1 \! - \! \beta^{2}_{d} \sin^{2}{\!\theta}\right]^{\!\frac{1}{2}} \! \cos^{2}{\!\theta} \, \mathrm{d}\theta,
\end{equation}
with $v_{\text{\tiny{F}}} \! \equiv \! \gamma / \hbar$ being referred to as Fermi velocity. Clearly, the NE electronic occupation, and  therefore the ``linear'' response of the out-of-equilibrium system outlined in this work, contains non-linear terms in the external field $\mathbf{E}_{\text{\tiny{DC}}}$. In what follows, we suppress this ``non-linearity'' and are thus able to use the standard linear-response Kubo formalism. In the low-current limit, i.e., $\beta_{d} \!\ll \! 1$, Eq. (\ref{Drift_Velocity}) yields $\mathrm{v}_{\! d} \! \cong \! \beta_{d} v_{\text{\tiny{F}}}$ and therefore a current density of $\mathrm{J} \! \cong \! \beta_{d} n_{s} v_{\text{\tiny{F}}}$. Unlike the experimental fashion in which the results would be reported in terms of the pump current density or the drain-source voltage, the numerical results in this work are presented in terms of the parameter $\beta_{d}$. 

Within the framework of SFD model, the application of drain-source voltage does not affect the energy dispersion of $\pi$ electrons, $\mathrm{E}^{s}_{\mathbf{k}}$, but leads to an anisotropic quasi Fermi energy $E_{\text{\tiny{F}}}(\theta_{\mathbf{k}},\theta_{d})$. In contrast, Refs. \citenum{Bistritzer,Svintsov12,Svintsov13,Serov} and \citenum{Duppen16} adopted an approach in which the NE distribution function is obtained through feeding Eq. (\ref{Fermi-Dirac}) with an isotropic Fermi \textit{level} $E_{\text{\tiny{F}}}$ while the energy eigen-states are given by the ones of tilted Dirac cones (TDC) \cite{Kobayashi,Georbig,KAWARABAYASHI,Yasuhiro,Trescher} whose energy dispersion is given by $\mathrm{E}^{s,\mathbf{k}}_{\text{tilt}} \equiv \mathrm{E}^{s}_{\mathbf{k}}- \hbar \mathbf{v}_{\! d} \cdot \mathbf{k}$. The electron density resulting from this approach would be dependent on temperature and drift velocity. Thus, for a given local drift velocity and temperature, the Fermi level $E_{\text{\tiny{F}}}$ should be adjusted to obtain the desired local electron density. In the limit of $\mathrm{v}_{\! d} \ll v_{\text{\tiny{F}}}$, the SFD and TDC models yield the same result. However, these models are not reliable within the large current regime, i.e., $\mathrm{v}_{\! d} \sim v_{\text{\tiny{F}}}$, and the BTE should be solved numerically \cite{PhysRevB.84.125450}.

\section{The optical response}\label{section:THE_OPTICAL_RESPONSE}
\subsection{Analytic and Semi-analytic approximations}\label{subsec:ANAP}
Applying the method introduced in Ref. \citenum{Behnaam15} to the conductivity integral given by Eq. (\ref{CONDUCTIVITY_TENSOR}) yields the following dyadic form for the optical conductivity tensor of the current-carrying $\pi$ electron gas (see Appendix \ref{Appendix:B}):
\begin{equation}\label{OPTICAL_CONDUCTIVITY_OF_CURRENT_CARRYING_ELECTRN_GAS_IN_DYADIC_FORM}
\tensor{\sigma}(\mathrm{q}\!=\!0,\omega) = \sigma_{\textsc{\tiny{L}}}(\omega) \, \hat{\mathbf{v}}_{\!d} \hat{\mathbf{v}}_{\!d} +  \sigma_{\textsc{\tiny{T}}}(\omega) \left[\tensor{\mathrm{I}} \! - \! \hat{\mathbf{v}}_{\!d} \hat{\mathbf{v}}_{\!d} \right],
\end{equation}
where $\hat{\mathbf{v}}_{\!d} \equiv \mathbf{v}_{\!d}/\mathrm{v}_{\! d}$ and the function $\sigma_{\textsc{\tiny{L/T}}}(\omega)$ is referred to as the longitudinal/transverse optical conductivity. At $T_{e}\!=\!0\hspace{0.08em}\mathrm{K}$, $\sigma_{\!\mu=\textsc{\tiny{L}},\textsc{\tiny{T}}}(\omega)$ can be obtained from:
\begin{equation}\label{LONGITUDINAL_AND_TRANSVERSE_COMPONENTS_OF_THE_CONDUCTIVITY_TENSOR}
\tilde{\sigma}_{\! \mu}(\omega) = i \frac{g_{s}g_{v}}{4 \pi^{2}} \sum_{\zeta=\pm} \int_{0}^{\pi} \!\! \mathrm{K}_{\zeta}^{\Gamma}\!(\omega,\theta) \left[1 \! + \! \lambda_{\mu}^{\zeta} \cos{\!(2\theta)}\right] \mathrm{d}\theta,
\end{equation}
where $g_{v} \! = \! 2$ is the valley degeneracy, $\tilde{\sigma}_{\! \mu} (\omega) \! \equiv \! \sigma_{\! \mu}(\omega)/\sigma_{\text{u}}$ and $\lambda_{\, \textsc{\tiny{L/T}}}^{\zeta} \! = \! \pm \zeta$. The kernel functions $\mathrm{K}_{\pm}^{\Gamma}\! (\omega,\theta)$ in Eq. (\ref{LONGITUDINAL_AND_TRANSVERSE_COMPONENTS_OF_THE_CONDUCTIVITY_TENSOR}) are expressed in terms of the nonequilibrium Fermi wavevector $k_{\scriptscriptstyle{F}}(\theta) \equiv \gamma^{-1}\left|E_{\scriptscriptstyle{F}}(\theta,\theta_{d}\!=\!0)\right|$ as follows:
\begin{eqnarray}
&\displaystyle{\mathrm{K}_{\scriptscriptstyle{+}}^{\Gamma}(\omega,\theta)\equiv \! 4 \frac{\gamma \, k_{\scriptscriptstyle{F}}(\theta) }{\hbar \omega \! + \! i \Gamma}} \label{INTRABAND_KERNEL_FUNCTION}
\\[1.0ex]
&\displaystyle{\mathrm{K}_{\scriptscriptstyle{-}}^{\Gamma}(\omega,\theta) \equiv \ln{ \! \left(\frac{2\gamma \, k_{\scriptscriptstyle{F}}(\theta) - [\hbar \omega \! + \! i \Gamma]}{2\gamma \, k_{\scriptscriptstyle{F}}(\theta) + [\hbar \omega \! + \! i \Gamma]}\right) }},\label{INTERBAND_KERNEL_FUNCTION}
\end{eqnarray}
where the role of $\Gamma$ is to take account of the disorder-induced scattering of $\pi$ electrons in a phenomenological manner \cite{PhysRevB.83.165113}. The equilibrium-state optical conductivity of graphene at $T_{e}\!=\!0K$ \cite{1367-2630-8-12-318,Falkovsky2007}, denoted here by $\tilde{\sigma}^{\0}(\omega)$, can be recovered from Eq. (\ref{LONGITUDINAL_AND_TRANSVERSE_COMPONENTS_OF_THE_CONDUCTIVITY_TENSOR}) in the $\beta_{d} \to 0$ limit,
\begin{equation}\label{EQUILIBRIUM_OPTICAL_CONDUCTIVITY}
\tilde{\sigma}^{\0}(\omega)\! = \! i \frac{g_{s}g_{v}}{4 \pi} \! \left[ \frac{4 \gamma k_{\scriptscriptstyle{F}}^{\0} }{\hbar \omega \! + \! i \Gamma} \! + \! \ln{ \! \left(\! \frac{2 \gamma k_{\scriptscriptstyle{F}}^{\0} \! - \! [\hbar \omega \! + \! i \Gamma]}{2 \gamma k_{\scriptscriptstyle{F}}^{\0} \! + \! [\hbar \omega \! + \! i \Gamma]}\right) } \right].
\end{equation}

The $\zeta\!=\!-1$ term in Eq. (\ref{LONGITUDINAL_AND_TRANSVERSE_COMPONENTS_OF_THE_CONDUCTIVITY_TENSOR}) corresponds to the interband optical conductivity whose intraband ($\zeta\!=\!+1$) counterpart is characterized as the ``Drude'' term, i.e., $\sigma^{\text{intra}} \! = \! \frac{i \hbar D}{\hbar \omega + i \Gamma}$, with the coefficient $D$, that is referred to as the Drude weight \cite{PhysRevB.83.165113,1367-2630-15-11-113050}, being altered in the presence of DC current. Thus, within the framework of SFD model, the longitudinal/transverse Drude weight of a current-carrying $\pi$ electron gas at $T_{e}\!=\!0\hspace{0.08em}\mathrm{K}$ is given by
\begin{equation}\label{GENERAL_DRUDE_WEIGHT}
D_{\textsc{\tiny{L/T}}} = \frac{D^{\0}}{\pi} \int_{0}^{\pi} \left[1 \! \pm \! \cos{\!(2\theta)}\right] \frac{k_{\text{\tiny{F}}}(\theta)}{k_{\text{\tiny{F}}}^{\0}} d\theta,
\end{equation}
with $D^{\0} \! = \!g_{s}g_{v} \sigma_{\!\text{u}} \frac{|E_{\text{\tiny{F}}}^{\0}|}{\pi\hbar}$ being the Drude weight at $\beta_{d}\!=\!0$ \cite{1367-2630-15-11-113050}. For large drift velocities, i.e., $\beta_d \! \to \! 1$, we have
\begin{equation}\label{LIMITING_CASE}
\frac{D_{\textsc{\tiny{L/T}}}}{D^{\0}} \! \to \! \frac{2}{\pi}\left[1 \pm \frac{1}{3}\right] \quad \Rightarrow \quad \frac{D_{\textsc{\tiny{L}}}}{D_{\textsc{\tiny{T}}}} \! \to \! 2,
\end{equation} 
which is analogous to the case of Black Phosphorus in which the anisotropic response can be largely attributed to the considerable difference between the logitudinal and transverse Drude weights \cite{PhysRevLett.113.106802}. The expansion of the Drude weight for small drift current
\begin{equation}\label{EXPANSION_OF_DRUDE_WEIGHT}
\frac{D_{\textsc{\tiny{L/T}}}}{D^{\0}} = 1-\frac{\beta_{d}^{2}}{4}\left[1 \! \mp \! \frac{1}{2}\right] + \mathcal{O}(\beta_{d}^{4}) ,
\end{equation}
indicates that the modification to the intraband optical conductivity is negligible in the low-current ($\beta_{d} \! \ll \! 1$) regime. The logarithmic divergence of the interband term, on the other hand, results in a pronounced modification within a frequency window centered at $\left| \omega \right| \! = \! 2 v_{\text{\tiny{F}}} k_{\text{\tiny{F}}}^{\0}$, in agreement with Pauli exclusion principle. At $T_{e}\!=\!0\hspace{0.08em}\mathrm{K}$, the real part of the low-frequency optical conductivity of driven $\pi$ electron gas in a clean sample of graphene, i.e., $\Gamma \! = \! 0$, is given by the following closed-form expression:
\begin{equation}\label{ANALYTIC_SOLUTION_FOR_THE_REAL_PART OF_CONDUCTIVITY}
\Re{[\tilde{\sigma}_{\textsc{\tiny{L/T}}}(\omega)]} =\frac{g_{s}g_{v}}{4\pi}\left[\vartheta \mp \cos{\vartheta} \left|\sin{\vartheta}\right|\right],
\end{equation}
with the angle $\vartheta$ being defined as (see Appendix \ref{Appendix:C}):
\begin{align}\label{VARTHETA_DEFINITION}
\vartheta \! \equiv \! \left\{
		\begin{array}{cc}
			 0 & \;\;\qquad \left| \tilde{\omega} \right| \leq  \tilde{\omega}_{\scriptscriptstyle{-}}
			\\[1.0ex]
			\!\!\! \arccos{\!\left[\frac{\tilde{\omega}_{\scriptscriptstyle{-}} \tilde{\omega}_{\scriptscriptstyle{+}} - \tilde{\omega}^{2}}{ 4 \left| \tilde{\omega} \right|  \beta_{d}}\right]}	&\qquad \tilde{\omega}_{\scriptscriptstyle{-}} \! \leq \left| \tilde{\omega} \right| \leq \tilde{\omega}_{\scriptscriptstyle{+}}
			\\[1.5ex]
			 \pi & \;\;\qquad \left| \tilde{\omega} \right| \geq \tilde{\omega}_{\scriptscriptstyle{+}}
		\end{array}
	\right.\!\!\!\!,
\end{align}
where $\tilde{\omega} \equiv \hbar \omega  / |E_{\text{\tiny{F}}}^{\0}|$ and $\tilde{\omega}_{\scriptscriptstyle{\pm}} \! \equiv 2 \left(1 \pm \beta_{d}\right)$. What is given by Eqs. (\ref{ANALYTIC_SOLUTION_FOR_THE_REAL_PART OF_CONDUCTIVITY}--\ref{VARTHETA_DEFINITION}) has a similar form to the one discussed in Ref. \citenum{Duppen16} wherein the choice of the NE distribution function obtained from the TDC results in an upper bound of the modification frequency range given by $\tilde{\omega}^{\text{\tiny{TDC}}}_{\scriptscriptstyle{+}} \equiv \frac{2}{1-(\mathrm{v}_{\! d}/v_{\text{\tiny{F}}})}$ which diverges in the $\mathrm{v}_{\! d} \! \to \! v_{\text{\tiny{F}}}$ limit.   

\subsection{The optical absorption spectra of current-carrying graphene}\label{subsec:OPT_ABS}
The absorption spectrum of current-carrying graphene, similar to the equilibrium-state measurements in Refs. \citenum{PhysRevLett.101.196405} and \citenum{PhysRevB.83.165113}, may provide evidence on the modification given by Eq. (\ref{ANALYTIC_SOLUTION_FOR_THE_REAL_PART OF_CONDUCTIVITY}). The absorption of a normally-incident EM plane wave by an anisotropic two-dimensional electron gas, sandwiched in between two dielectrics, $A$, is formulated as follows (see Appendix \ref{Appendix:D}):
\begin{equation}\label{OPTICAL_ABSORPTION_OF_ANISOTROPIC_2D_ELECTRON_GAS}
A =  4 \pi \alpha \frac{\Re{[\tilde{\sigma}_{\textsc{\tiny{L}}}]} \cos^{2}{\! \phi } + \Re{[\tilde{\sigma}_{\textsc{\tiny{T}}}]} \sin^{2}{\! \phi}} {\sqrt{\frac{\mu^{r}_{1}}{\varepsilon^{r}_{1}}}\left[\sqrt{\frac{\varepsilon^{r}_{1}}{\mu^{r}_{1}}} + \sqrt{\frac{\varepsilon^{r}_{2}}{\mu^{r}_{2}}} \, \right]^{2}} +  \mathcal{O}(\alpha^{2}),
\end{equation}
with $\alpha \equiv \frac{e^{2}}{4\pi \varepsilon_{\textsc{\tiny{0}}} \hbar c}$, $\varepsilon_{\textsc{\tiny{0}}}$, $c$, $\phi \equiv \theta_{\mathbf{i}}^{p} - \theta_{d}$, and $\varepsilon^{r}_{j}$ ($\mu^{r}_{j}$) being the fine structure constant, the permittivity of vacuum, the phase velocity of light in vacuum, the angle between the polarization of the normally-incident EM wave and the drift velocity, and the relative permittivity (permeability) of the $j$-th optical medium. The expression given by Eq. (\ref{OPTICAL_ABSORPTION_OF_ANISOTROPIC_2D_ELECTRON_GAS}) describes the case in which the transmitted (reflected) EM wave propagates through medium $2$ ($1$).
\begin{figure}
	\begin{center}
		\includegraphics[clip,width=0.98\columnwidth]{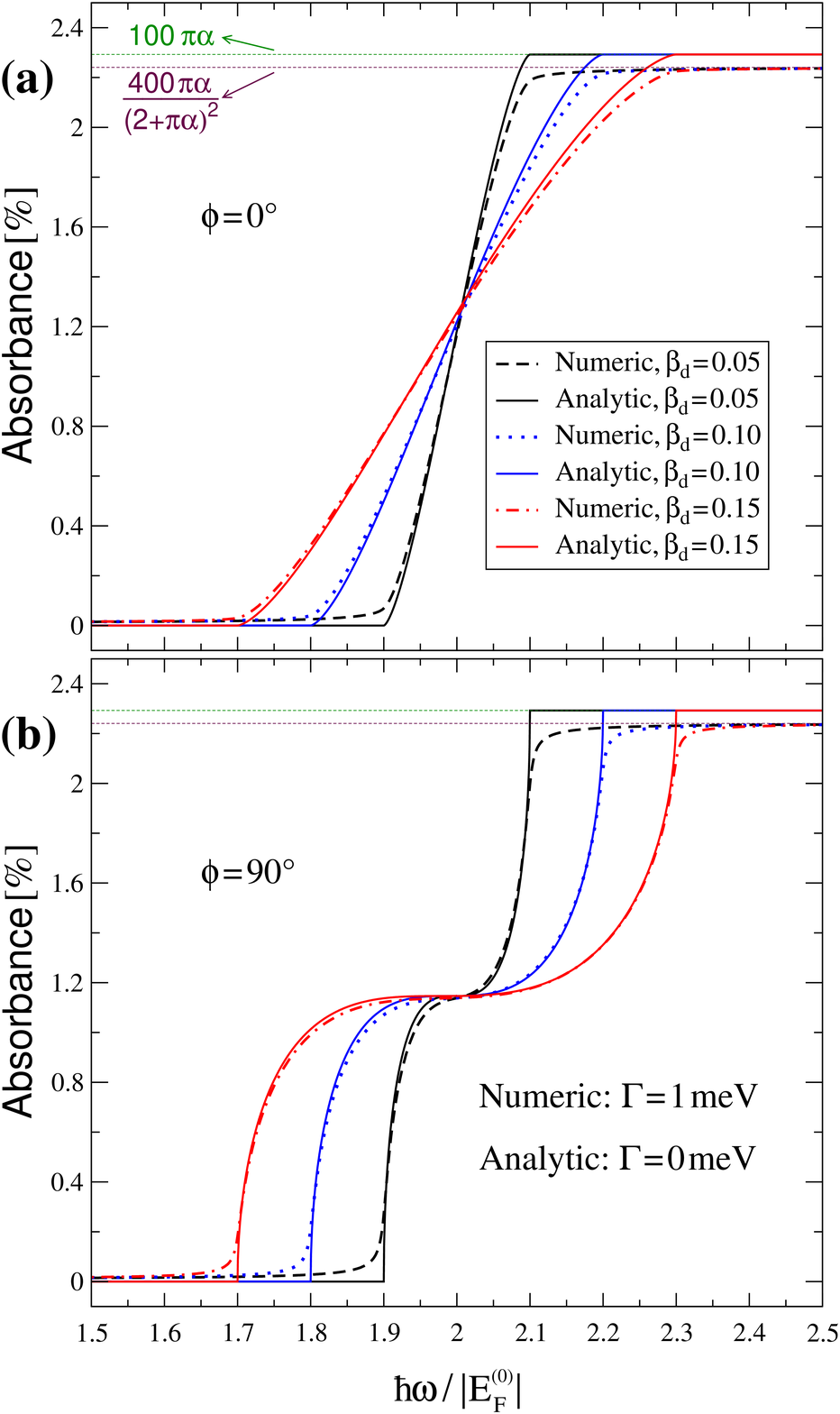}\\
		\caption{(Color online) The absorption spectra of a suspended (i.e. $\varepsilon^{r}_{1,2} \! = \! \mu^{r}_{1,2} \! = \! 1$) current-carrying graphene sample at $T_{e}\! = \!0K$ as simulated by the SFD model for the two cases wherein the polarization of the normally incident EM wave is \textbf{(a)} parallel and \textbf{(b)} perpendicular to the drift velocity. The curves labeled as ``Numeric'' are generated via plugging the output of Eq. (\ref{LONGITUDINAL_AND_TRANSVERSE_COMPONENTS_OF_THE_CONDUCTIVITY_TENSOR}) into the Fresnel's coefficients, while their ``Analytic'' counterparts are the product of Eqs.  (\ref{ANALYTIC_SOLUTION_FOR_THE_REAL_PART OF_CONDUCTIVITY}--\ref{OPTICAL_ABSORPTION_OF_ANISOTROPIC_2D_ELECTRON_GAS}).}
		\label{OPTICAL_ABSORPTION_OF_CURRENT_CARRYING_GRAPHENE}
	\end{center}
\end{figure}
\subsection{Current-induced Kerr and Faraday rotations}\label{subsec:KERR_FARADAY}
Aside from its impact on the absorption spectra, the DC electric current converts the linear polarization of the incident (\textit{i}) EM wave into elliptic for the reflected (\textit{r}) and transmitted (\textit{t}) EM waves  \cite{PhysRevB.81.115413}. The polarization angle of the \textit{r}-wave (\textit{t}-wave), denoted here by $\theta^{p}_{r}$ ($\theta^{p}_{t}$), is defined as the angle between the $x$-axis and the major axis of the ellipse that is being traced out by the tip of the electric-field vector of the \textit{r}-wave (\textit{t}-wave). Due to the current-induced birefringence, the polarization of the \textit{r}-wave (\textit{t}-wave) gets rotated with respect to that of the \textit{i}-wave: a phenomenon known as Kerr (Faraday) rotation that is quantitatively described via defining the Kerr (Faraday) rotation angle as $\theta_{\text{\tiny{K}}} \! \equiv \! \theta^{p}_{r} \! - \! \theta^{p}_{\mathbf{i}}$ ($\theta_{\text{\tiny{F}}} \! \equiv \! \theta^{p}_{t} \! - \!
  \theta^{p}_{\mathbf{i}}$). The Kerr rotation angle is given by (see Appendix \ref{Appendix:E}):
\begin{equation}\label{KERR_ANGLE}
\tan{[2\theta_{\text{\tiny{K}}}]}=\frac{\tan{[2\xi_{r}^{x}]} \cos{\psi_{r}}-\tan{[2\theta^{p}_{\mathbf{i}}]}}{1+\tan{[2\theta^{p}_{\mathbf{i}}]}\tan{[2\xi_{r}^{x}]} \cos{\psi_{r}}},
\end{equation}
where $\psi_{r} \equiv \arg{\!\left( \!\textstyle{\frac{r_{xx}}{r_{yy}}}\!\right)}$ and $\tan{\xi_{r}^{x}} \equiv \frac{|r_{yy}|}{|r_{xx}|} \tan{\theta_{i}^{p}}$ with $r_{xx}$ and $r_{yy}$ being the Fresnel reflection coefficients,
\begin{equation}\label{FRESNEL_REFLECTION_COEFFICIENTS_NORMAL_INCIDENCE_XX_AND_YY}
r_{xx(yy)}=\frac{\sqrt{\frac{\varepsilon^{r}_{1}}{\mu^{r}_{1}}}-\sqrt{\frac{\varepsilon^{r}_{2}}{\mu^{r}_{2}}}-\pi\alpha \, \tilde{\sigma}_{\textsc{\tiny{L(T)}}}} {\sqrt{\frac{\varepsilon^{r}_{1}}{\mu^{r}_{1}}}+\sqrt{\frac{\varepsilon^{r}_{2}}{\mu^{r}_{2}}}+\pi\alpha \, \tilde{\sigma}_{\textsc{\tiny{L(T)}}}}.
\end{equation}
In addition, the Faraday rotation angle $\theta_{\text{\tiny{F}}}$ can be obtained via replacing the Fresnel reflection coefficients $r_{nn}$ ($n=x,y$) in Eq. (\ref{KERR_ANGLE}) with the Fresnel transmission coefficients $t_{nn}=1+r_{nn}$ (see Appendix \ref{Appendix:E}). As presented in Fig. \ref{KERR_ROTATION}-\textbf{(a)}, large Kerr rotation angles ($\theta_{\text{\tiny{K}}} \! \sim \! 10^{\circ}$) can be achieved with suspended graphene. The extremely small reflectance ($R \! \sim \! 10^{-4}$), however, hinders the observation of such large Kerr rotation angles. Nonetheless, a small enough difference between the permittivities of the surrounding dielectrics is expected to yield large Kerr rotation while the increased reflectance allows for measurements. On the other hand, Fig. \ref{KERR_ROTATION}-\textbf{(b)} presents small Kerr rotation angles $\theta_{\text{\tiny{K}}} \! \sim \! 0.1^{\circ}$ for the case of graphene lying on a hexagonal boron nitride (hBN) \cite{Dean,LeRoy,Woe14} substrate, while the reflectance is large ($R \! \sim \! 10 \%$). This indicates the high sensitivity of the Kerr rotation to the choice of the top and bottom dielectrics. However, as presented in Fig. \ref{FARADAY_ROTATION}, the Faraday rotation does not exhibit such sensitivity.
\begin{figure}
	\begin{center}
		\includegraphics[clip,width=0.98\columnwidth]{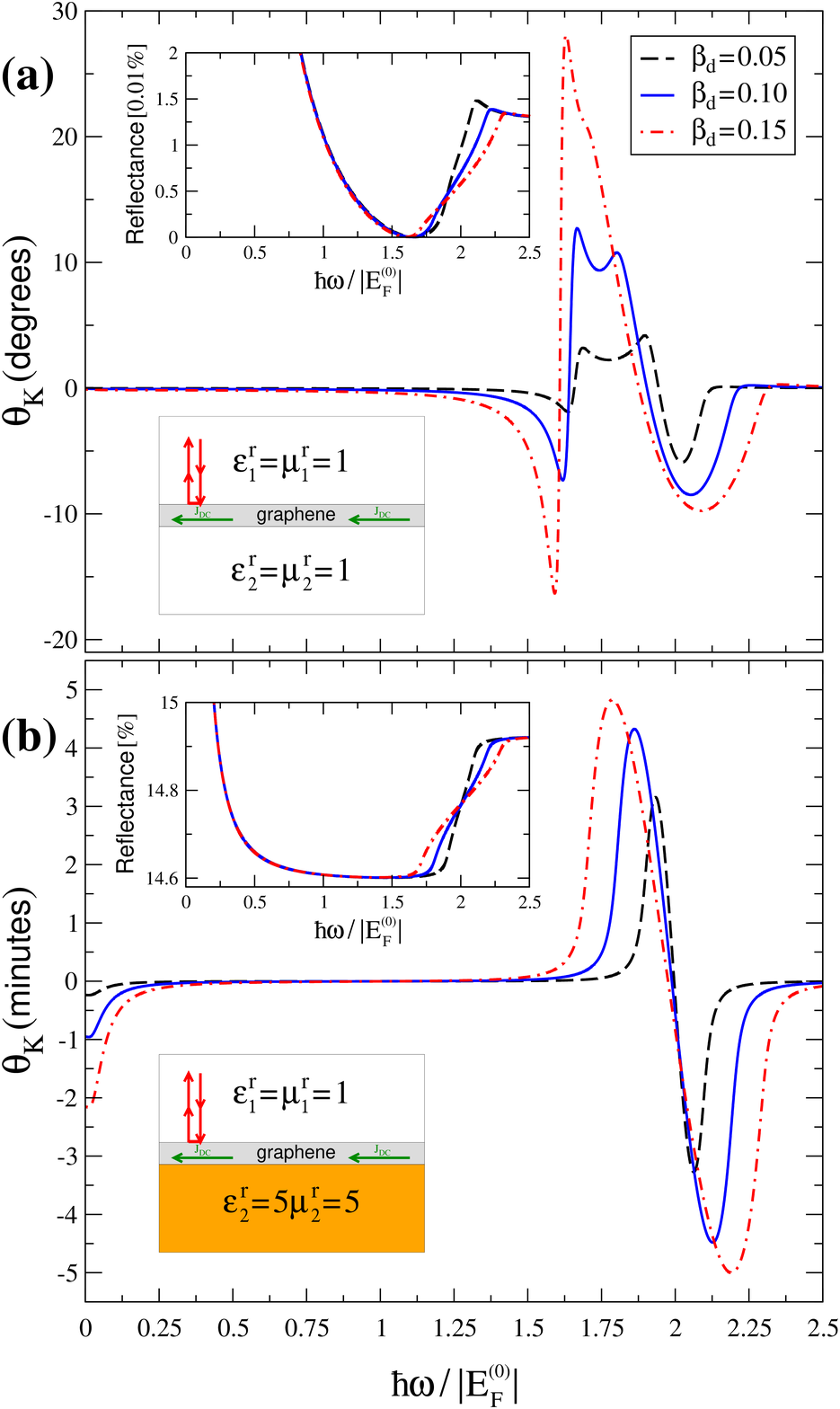}\\
		\caption{(Color online) The Kerr rotation data obtained from Eqs. (\ref{LONGITUDINAL_AND_TRANSVERSE_COMPONENTS_OF_THE_CONDUCTIVITY_TENSOR}) and (\ref{KERR_ANGLE}) for a current-carrying $\pi$ electron gas at $T_{e}\!=\!0\hspace{0.08em}\mathrm{K}$ being contained by a disordered ($\Gamma \! = \! 5 \, meV$) sample of \textbf{(a)} suspended and \textbf{(b)} on-substrate graphene. Both panels present the maximal Kerr data corresponding to the geometry wherein the polarization of the normally-incident EM wave makes an angle of $\phi \equiv \theta_{\mathbf{i}}^{p} \! -  \theta_{d} \! = \! 45^{\circ}$ with the drift velocity $\mathbf{v}_{\!d}$. The Kerr rotations in the upper/lower panel are reported in degrees/minutes.}
		\label{KERR_ROTATION}
	\end{center}
\end{figure}
\begin{figure}	
	\begin{center}
		\includegraphics[clip,width=0.98\columnwidth ]{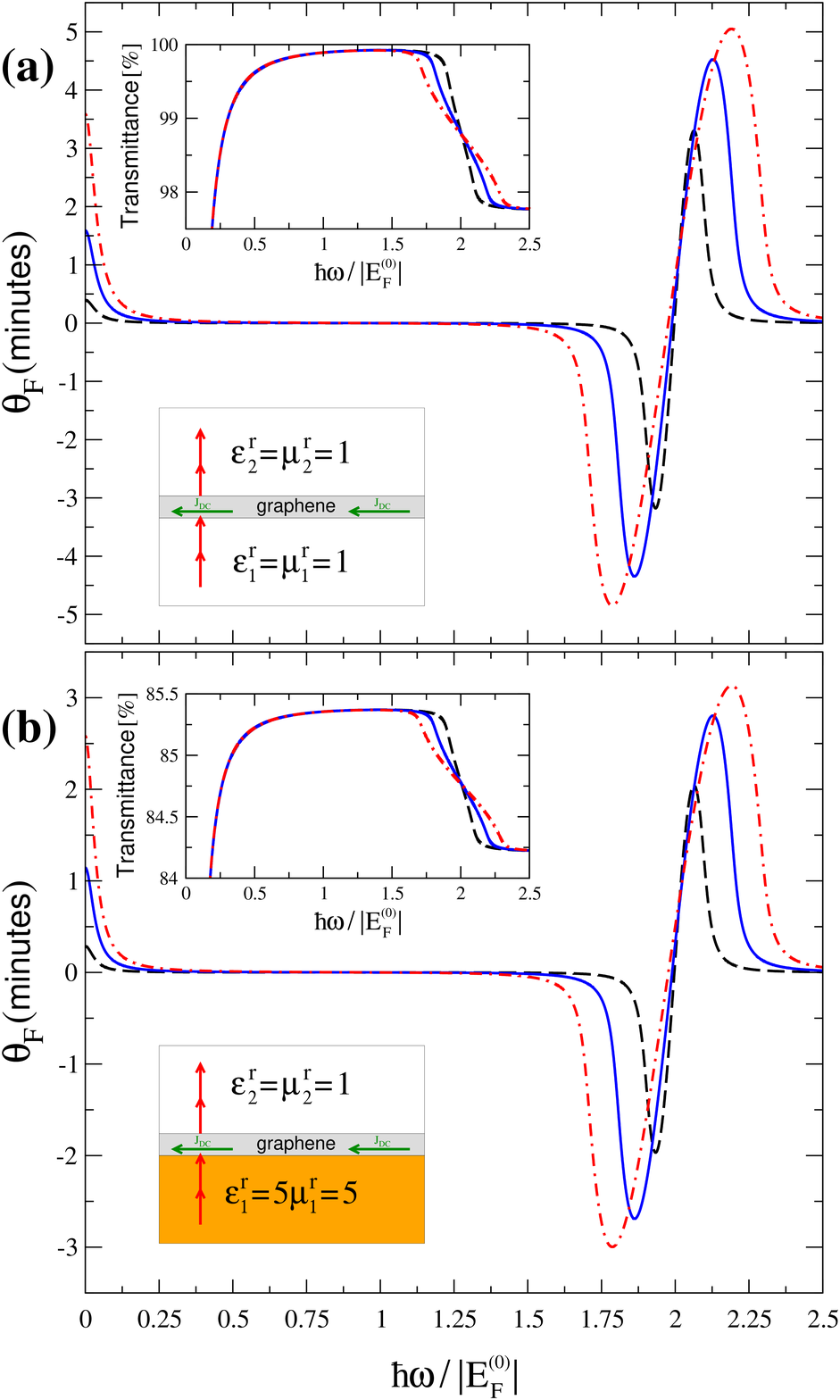}\\
		\caption{(Color online) Faraday rotation data with the same descriptions as those provided for Fig. \ref{KERR_ROTATION}.}
		\label{FARADAY_ROTATION}
	\end{center}
\end{figure}

As shown in Fig. \ref{FARADAY_ROTATION}, the Faraday rotation achieved through a single transmission is small; however, different experimental techniques could be implemented to obtain larger rotation angles through multiple transmissions. Our calculations indicate that a configuration in which graphene is sandwiched in between hBN dielectrics, i.e., $\varepsilon^{r}_{1,2} \! = \! 5 \mu^{r}_{1,2} \! = \! 5$, yields a minimum transmittance of $99\%$ which allows for trading off the total transmittance for a larger Faraday angle.

The disorder-induced scattering of $\pi$ electrons, the Joule heating of current-carrying sample \cite{HEINZ,KWKIM,BAE_POP,KOZ} and the electron density fluctuation \cite{Yacoby8,LeRoy,Decker} in the area under the probe EM beam are the major factors which cause the smoothing of the logarithmic divergence of the interband optical conductivity, and therefore, hinder the observation of the electro-optical Kerr/Faraday rotations discussed here. The high precision of $10^{-5}\hspace{0.08em}\mathrm{rad}$ with which the Kerr rotations were measured in Ref. \citenum{Jieun} suggests that the rotations reported here, though as tiny as $1' \! \approx 291 \! \times \! 10^{-6}\hspace{0.08em}\mathrm{rad}$, are measurable. Since large amounts of DC current are expected to cause the longitudinal and transverse Drude weights to be considerably different, even for the case of identical surrounding dielectrics, the low-frequency electro-optical Kerr/Faraday rotations cannot be neglected out of the low-current regime.

\section{Conclusions}\label{section:SC}

We theoretically discussed the non-equilibrium (NE) response of a monolayer graphene that carries a DC current. Our analytical and numerical calculation results indicate that a DC-current-carrying monolayer graphene can exhibit larger Kerr rotations. For this calculation, we employ the phenomenological shifted Fermi disk (SFD) model. Future works may study the NE response of current-carrying graphene via feeding the conductivity integral [Eq. (\ref{CONDUCTIVITY_TENSOR})] with the NE electronic occupation obtained from numerical BTE solvers.

Discussing the optical response, we find that local measurements of the Kerr/Faraday rotation angle or absorbance, within a tunable frequency window, can be exploited to determine the current density distribution throughout the whole channel. More specifically, this can be achieved through generating 2D Kerr maps of the current-carrying channel, as in Ref. \citenum{Jieun}, and then converting the Kerr maps into current maps.

The numerical estimates presented in this work are specific to $T\!=\!0\hspace{0.08em}\mathrm{K}$ and are based on the SFD model, and therefore, are valid within the low-current regime. Nonetheless, the electro-optical phenomena discussed here are expected to be observable within the high-current regime, i.e., $\mathrm{v}_{\!d} \! \sim \! v_{\text{\tiny{F}}}$, provided that the tilt of the Fermi ``level'' induced by the drain-source voltage is larger than the thermal fluctuations, i.e., $\gamma \mathrm{k}_{\text{shift}} \! \gg \! k_{\text{\tiny{B}}}T_{e}$. In this case, the experimental measurements are expected to be in qualitative agreement with this work, but require a more realistic modeling to be numerically reproduced.   

\begin{acknowledgments}

This research was funded by the National Research Foundation of Korea (Science Research Center Program, 2011-0030046) and Spain's Ministerio de Econom{\'i}a, Industria y Competitividad (FIS2017-82260-P, FIS2014-57432-P). M.S. would like to express his sincere gratitude towards his supervisor at POSTECH, Prof. Kwang S. Kim, for his support and valuable advice, and towards the members and staff of ICMM (Instituto de Ciencia de Materiales de Madrid) for their hospitality.

\end{acknowledgments}
\appendix
\renewcommand{\theequation}{A\arabic{equation}}
\setcounter{equation}{0}
\section{The drift velocity within the framework of the SFD model}\label{Appendix:A}
The following summation defines the drift velocity: 
\begin{equation}\label{DRIFT_VELOCITY_DEFINITION}
\mathbf{v}_{\!d} \equiv \frac{g_{s}}{N_{e}}\sum_{\mathbf{k} \in \scriptscriptstyle{F} \! \scriptscriptstyle{B} \! \scriptscriptstyle{Z}} { \sum_{s=\pm} {n_{\scriptscriptstyle{F}}[E_{s}(\mathbf{k})] v_{g}^{s}(\mathbf{k}) }},
\end{equation}
where $N_{e}$ is the number of the electrons brought into ($E^{\0}_{\scriptscriptstyle{F}}\!>\!0$) or taken out ($E^{\0}_{\scriptscriptstyle{F}}\!<\!0$) of graphene via doping, and $v_{g}^{s}(\mathbf{k})$ is the semi-classically-defined group velocity  corresponding to the eigen-state $\ket{\mathbf{k},s}$ given by
\begin{equation}\label{GROUP_VELOCITY_DEFINITION}
v_{g}^{s}(\mathbf{k}) \equiv \frac{1}{\hbar} \nabla_{\!\mathbf{k}}E_{s}(\mathbf{k}).
\end{equation}
The only contribution to the summation in Eq. (\ref{DRIFT_VELOCITY_DEFINITION}) is usually made by the eigen-states near the Dirac points; thus, the linear energy dispersion $E_{s}(\mathbf{k}) \cong s \gamma \mathrm{k}$ should be sufficient. As a result, the group velocity reads as
\begin{equation}\label{GROUP_VELOCITY}
v_{g}^{s}(\mathbf{k}) \cong s \frac{\gamma}{\hbar} \left[ \hat{\mathbf{\mathbf{e}}}_{x} \cos{\theta_{\mathbf{k}}} + \hat{\mathbf{\mathbf{e}}}_{y} \sin{\theta_{\mathbf{k}}} \right] =s v_{\scriptscriptstyle{F}} \hat{\mathbf{k}},
\end{equation}
and, the definition given by Eq. (\ref{DRIFT_VELOCITY_DEFINITION}) evolves into
\begin{equation}\label{DRIFT_VELOCITY_INTEGRAL_1}
\frac{\mathbf{v}_{\!d}}{v_{\scriptscriptstyle{F}}} = \frac{g_{s} g_{v}} {\left(2\pi\right)^{2}} \frac{\text{sgn}[E_{\scriptscriptstyle{F}}^{\0}]}{n_{s}}\int_{0}^{2\pi} \hat{\mathbf{k}} \, \mathrm{d}\theta_{\mathbf{k}} \int_{0}^{k_{\scriptscriptstyle{F}}(\theta_{\mathbf{k}},\theta_{d})}   \mathrm{k}\, \mathrm{d}\mathrm{k}.
\end{equation}
Since $n_{s} = [k_{\scriptscriptstyle{F}}^{\0}]^{2}/\pi$, the preceding integral becomes
\begin{equation}\label{DRIFT_VELOCITY_INTEGRAL_2}
\frac{\mathbf{v}_{\!d}}{v_{\scriptscriptstyle{F}}}=\frac{\text{sgn}[E_{\scriptscriptstyle{F}}^{\0}]}{2 \pi} \int_{0}^{2\pi} \left[\frac{E_{\scriptscriptstyle{F}}(\theta_{\mathbf{k}},\theta_{d})}{E_{\scriptscriptstyle{F}}^{\0}}\right]^{2} \hat{\mathbf{k}} \, \mathrm{d}\theta_{\mathbf{k}}.
\end{equation}
Plugging the expression given by Eq. (\ref{Nonequilibrium_EF}) into Eq. (\ref{DRIFT_VELOCITY_INTEGRAL_2}) is the last step in obtaining what presented by Eq. (\ref{Drift_Velocity}).
\begin{figure}
	\begin{center}
		\includegraphics[clip,width=0.98\columnwidth]{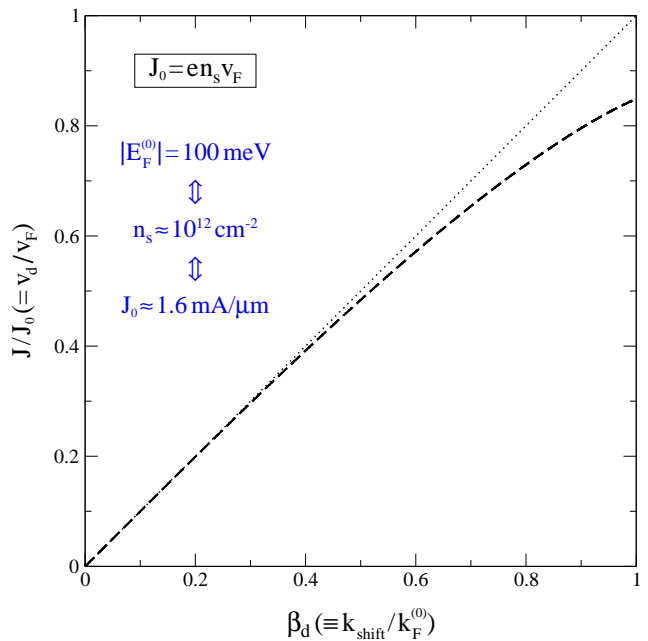}\\
		\caption{(Color online) Surface current density simulated by the SFD model [obtained from Eq. (\ref{Drift_Velocity})] for the whole range of the shift parameter, i.e. $0\leq\beta_{d}\leq1$. The order of magnitude of the simulated current density is in agreement with other numerical simulations \cite{Bae,PhysRevB.84.125450,Serov} as well as the experimental data \cite{4800722,Schwierz,PhysRevLett.103.076601,Bae} within the low-current regime.}
		\label{CURRENT_DENSITY}
	\end{center}
\end{figure}
%
\renewcommand{\theequation}{B\arabic{equation}}
\setcounter{equation}{0}
\section{Semi-analytic expression for the optical conductivity of current-carrying $\pi$ electron gas at $T_{e}\!=\!0K$}\label{Appendix:B}
The interband ($s\overline{s}\!=\!-1$) optical ($\mathrm{q}\!=\!0$) conductivity given by the single-valley ($g_{v}\!=\!2$) form of Eq. (\ref{CONDUCTIVITY_TENSOR}) can be arranged into the following dimensionless expression \cite{Behnaam15}:
\begin{equation}\label{INTERBAND_OPTICAL_CONDUCTIVITY}
\tilde{\sigma}_{n,\overline{n}}^{\text{inter}}(\omega) = \tilde{\sigma}_{n,\overline{n}}^{\scriptscriptstyle{E}_{\scriptscriptstyle{F}}\scriptscriptstyle{=0}}(\omega) + i \frac{g_{s}g_{v}}{4\pi^{2}} \! \int_{0}^{2\pi} {\!\!\! \Lambda_{n,\overline{n}}(\omega,\theta_{\mathbf{k}}) \, \mathrm{d}\theta_{\mathbf{k}}},
\end{equation}
with $\tilde{\sigma}_{n,\overline{n}}^{\text{inter}}\! \equiv \! \sigma_{n,\overline{n}}^{\text{inter}}/[e^{2}/(4 \hbar)]$ denoting the dimensionless interband optical conductivity of the $\pi$ electron gas in graphene at $T_{e}\!=\!0K$ and $\Lambda_{n,\overline{n}}(\omega,\theta_{\mathbf{k}})$ defined through
\begin{equation}\label{DEFINITION_OF_THE_POLAR_INTEGRAND}
\Lambda_{n,\overline{n}}(\omega,\theta_{\mathbf{k}}) \equiv \mathrm{F}_{\!n,\overline{n}}^{-}\! \left(\theta_{\mathbf{k}}\right) \ln{ \! \left( \! \frac{2|E_{\scriptscriptstyle{F}}(\theta_{\mathbf{k}},\theta_{d})| \! - \! \hbar \omega^{\prime}}{2|E_{\scriptscriptstyle{F}}(\theta_{\mathbf{k}},\theta_{d})| \! + \! \hbar \omega^{\prime}} \! \right) },
\end{equation}
where $\omega^{\prime} \equiv \omega+i0^{\scriptscriptstyle{+}}$ and $\mathrm{F}_{\!n,\overline{n}}^{s \overline{s}}\!\left(\theta_{\mathbf{k}}\right)$ denotes the optical limit ($\mathrm{q}\!=\!0$) of the expressions presented by Eqs. (\ref{BAND_OVERLAP_INTEGRAL_XX}--\ref{BAND_OVERLAP_INTEGRAL_YY}):
\begin{equation}\label{OPTICAL_BAND_OVERLAP_INTEGRAL_XX}
2 \, \mathrm{F}_{\!x,x}^{s \overline{s}} \! \left(\theta_{\mathbf{k}}\right) \cong 1 + s\overline{s} \, \cos{[2\theta_{\mathbf{k}}]}\,
\end{equation}
\begin{equation}\label{OPTICAL_BAND_OVERLAP_INTEGRAL_XY}
2 \, \mathrm{F}_{\!x,y}^{s \overline{s}} \! \left(\theta_{\mathbf{k}}\right) = 2 \, \mathrm{F}_{\!y,x}^{s \overline{s}} \! \left(\theta_{\mathbf{k}}\right) \cong s\overline{s} \, \sin{[2\theta_{\mathbf{k}}]}
\end{equation}
\begin{equation}\label{OPTICAL_BAND_OVERLAP_INTEGRAL_YY}
2 \, \mathrm{F}_{\!y,y}^{s \overline{s}} \! \left(\theta_{\mathbf{k}}\right) \cong 1 - s\overline{s} \, \cos{[2\theta_{\mathbf{k}}]}.
\end{equation}
On the other hand, at $T_{e}\!=\!0K$, the intraband ($s\overline{s}\!=\!+1$) optical ($\mathrm{q}\!=\!0$) conductivity given by the single-valley ($g_{v}\!=\!2$) form of Eq. (\ref{CONDUCTIVITY_TENSOR}) can be expressed as follows \cite{Falkovsky2007}:
\begin{equation}\label{INTRABAND_OPTICAL_CONDUCTIVITY}
\tilde{\sigma}_{n,\overline{n}}^{\text{intra}}(\omega) = i \frac{g_{s}g_{v}}{\pi^{2}\hbar \omega^{\prime}}\! \int_{0}^{2\pi} {\!\!\! \mathrm{F}_{\!n,\overline{n}}^{+} \! \left(\theta_{\mathbf{k}}\right) |E_{\scriptscriptstyle{F}}(\theta_{\mathbf{k}},\theta_{d})| \, \mathrm{d}\theta_{\mathbf{k}}}.
\end{equation}
The last step is to take the dependence on $\theta_{d}$ out of the integrals in Eqs. (\ref{INTERBAND_OPTICAL_CONDUCTIVITY}) and (\ref{INTRABAND_OPTICAL_CONDUCTIVITY}) through changing the integration variable into $\theta \equiv  \theta_{\mathbf{k}} - \theta_{d}$. As a result, the $\sin{\theta_{d}}$ and $\cos{\theta_{d}}$ multipliers emerging from $\mathrm{F}_{\!n,\overline{n}}^{s \overline{s}}(\theta_{\mathbf{k}})$ can be moved out of the integral. Of course, the terms in the conductivity integrals corresponding to the band overlap of $\sin{\left(2\theta\right)}$ vanish. Moreover, the $f(\theta)=f(-\theta)$ symmetry exhibited by the integrands for $0 \leq \theta \leq \pi$ allows us to reduce the range of integration. Rewriting the final result in terms of drift velocity leads us to the expressions presented by Eqs. (\ref{OPTICAL_CONDUCTIVITY_OF_CURRENT_CARRYING_ELECTRN_GAS_IN_DYADIC_FORM}--\ref{INTERBAND_KERNEL_FUNCTION}).
\renewcommand{\theequation}{C\arabic{equation}}
\setcounter{equation}{0}
\section{Analytic expression for the real part of the optical conductivity of current-carrying $\pi$ electron gas at $T_{e}\!=\!0K$}\label{Appendix:C}
The real part of the intraband optical conductivity given by Eq. (\ref{LONGITUDINAL_AND_TRANSVERSE_COMPONENTS_OF_THE_CONDUCTIVITY_TENSOR}) of a \textit{clean} ($\Gamma\!=\!0$) sample of graphene reduces to the Dirac delta function, $\delta_{\!\scriptscriptstyle{D}}(\omega)$, i.e.
\begin{equation}\label{INTERBAND_OPTICAL_CONDUCTIVITY_OF_A_CLEAN_SAMPLE}
\lim_{\Gamma \to 0} \Re{\left[\tilde{\sigma}_{\! \mu}^{\text{intra}}(\omega)\right]} = \pi D_{\! \mu} \, \delta_{\!\scriptscriptstyle{D}}(\omega),
\end{equation}
which means the optical conductivity of a clean sample of graphene is merely due to the interband transitions. For brevity, we define the function $\mathrm{M}^{\Gamma}\!(\omega,\theta)$ to be
\begin{equation}\label{THE_SPACIOUS_FRACTION}
\mathrm{M}^{\Gamma}\!(\omega,\theta) \equiv \frac{2\gamma \, k_{\scriptscriptstyle{F}}(\theta) - [\hbar \omega \! + \! i \Gamma]}{2\gamma \, k_{\scriptscriptstyle{F}}(\theta) + [\hbar \omega \! + \! i \Gamma]}.
\end{equation}
Combining the identities given by $\Im{\left[\ln{\!\left(z\right)}\right]} \! = \! \arg{\!\left(z\right)}$ and $\arg{\!\left(-r\right)} \! = \! - \pi \, \Theta \!\left[ r \right]$ (where $\Theta$, $z$ and $r$ respectively denote the Heaviside step function, a complex number, and a real number) yields
\begin{equation}\label{CLEAN_SAMPLE_LIMIT_OF_THE_ARGUMENT}
\lim_{\Gamma \to 0}{\Im{\left[\ln{\!\left(\mathrm{M}^{\Gamma}\!(\omega,\theta)\right)}\right]}} = - \pi \, \Theta \!\left[ \hbar \left|\omega\right| - 2 \gamma \, k_{\scriptscriptstyle{F}}(\theta)\right].
\end{equation}
Applying the preceding relation to the real part of the interband optical conductivity given by Eq. (\ref{LONGITUDINAL_AND_TRANSVERSE_COMPONENTS_OF_THE_CONDUCTIVITY_TENSOR}) returns a simplified expression in the clean-sample limit:
\begin{equation}\label{SIMPLIFIED_EXPRESSION_FOR_THE_REAL_PART_OF_CONDUCTIVITY}
\Re{\left[\tilde{\sigma}_{\textsc{\tiny{L/T}}}(\omega)\right]} = \frac{g_{s}g_{v}}{4 \pi} \int_{\pi-\vartheta}^{\pi} \left[1\mp \cos{\left(2 \theta \right)}\right] \, \mathrm{d}\theta,
\end{equation}
where the angle $0 \leq \vartheta \leq \pi$, as described by Eq. (\ref{VARTHETA_DEFINITION}), can be obtained through searching for the solutions of $2|E_{\scriptscriptstyle{F}}(\pi-\vartheta,0)| \! = \! \hbar \! \left|\omega\right|$. Evaluating the integral in Eq. (\ref{SIMPLIFIED_EXPRESSION_FOR_THE_REAL_PART_OF_CONDUCTIVITY}) is the last step to the expression given by Eq. (\ref{ANALYTIC_SOLUTION_FOR_THE_REAL_PART OF_CONDUCTIVITY}).
\renewcommand{\theequation}{D\arabic{equation}}
\setcounter{equation}{0}
\section{The optical absorption of an anisotropic two-dimensional electron gas}\label{Appendix:D}
The flux of EM energy associated with each of the normally-incident (\textit{i}), reflected (\textit{r}) and transmitted (\textit{t}) plane waves is given by their respective time-averaged Poynting vectors ($\overline{\mathbf{\mathbf{E}}}_{w=i,t,r}$ is the amplitude of \textit{w}-wave):
\begin{equation}\label{TIME_AVERAGED_POYNTING_VECTOR_OF_THE_INCIDENT_EM_WAVE}
\langle\mathbf{S}_{i}\rangle=-\frac{1}{2} \sqrt{\frac{\varepsilon^{r}_{1} \varepsilon_{\textsc{\tiny{0}}}}{ \mu^{r}_{1}\mu_{\textsc{\tiny{0}}}}} \left[\overline{\mathbf{\mathbf{E}}}_{i} \cdot \overline{\mathbf{\mathbf{E}}}^{*}_{i}\right] \hat{\mathbf{\mathbf{e}}}_{z}\,
\end{equation}
\begin{equation}\label{TIME_AVERAGED_POYNTING_VECTOR_OF_THE_REFLECTED_EM_WAVE}
\langle\mathbf{S}_{r}\rangle=+\frac{1}{2} \sqrt{\frac{\varepsilon^{r}_{1} \varepsilon_{\textsc{\tiny{0}}}}{ \mu^{r}_{1}\mu_{\textsc{\tiny{0}}}}}\left[\overline{\mathbf{\mathbf{E}}}_{r} \cdot \overline{\mathbf{\mathbf{E}}}^{*}_{r}\right] \hat{\mathbf{\mathbf{e}}}_{z}\,
\end{equation}
\begin{equation}\label{TIME_AVERAGED_POYNTING_VECTOR_OF_THE_TRANSMITTED_EM_WAVE}
\langle\mathbf{S}_{t}\rangle=-\frac{1}{2} \sqrt{\frac{\varepsilon^{r}_{2} \varepsilon_{\textsc{\tiny{0}}}}{ \mu^{r}_{2}\mu_{\textsc{\tiny{0}}}}}\left[\overline{\mathbf{\mathbf{E}}}_{t} \cdot \overline{\mathbf{\mathbf{E}}}^{*}_{t}\right] \hat{\mathbf{\mathbf{e}}}_{z}.
\end{equation}
If the Cartesian coordinate system is positioned so that the $x$-axis is aligned with the drift velocity, i.e. $\mathbf{v}_{\!d}\!=\! \mathrm{v}_{\! d} \hat{\mathbf{\mathbf{e}}}_{x}$, Eq. (\ref{OPTICAL_CONDUCTIVITY_OF_CURRENT_CARRYING_ELECTRN_GAS_IN_DYADIC_FORM}) yields a diagonal optical conductivity tensor and, in consequence, the off-diagonal Fresnel reflection coefficients vanish, i.e. $r_{xy}\!=\!r_{yx}\!=\!0$. In this case, the amplitude-vector of the \textit{r} and \textit{t} waves can be expressed in terms of the amplitude and polarization angle of \textit{i}-wave:
\begin{equation}\label{THE_AMPLITUDE_OF_THE_REFLECTED_EM_WAVE}
\overline{\mathbf{\mathbf{E}}}_{r}=\left|\overline{\mathbf{\mathbf{E}}}_{i}\right| \left \{r_{xx} \cos{\theta_{i}^{p}}\,\hat{\mathbf{\mathbf{e}}}_{x} + r_{yy} \sin{\theta_{i}^{p}} \,\hat{\mathbf{\mathbf{e}}}_{y} \right \}\,
\end{equation}
\begin{equation}\label{THE_AMPLITUDE_OF_THE_TRANSMITTED_EM_WAVE}
\overline{\mathbf{\mathbf{E}}}_{t}=\left|\overline{\mathbf{\mathbf{E}}}_{i}\right| \left \{t_{xx} \cos{\theta_{i}^{p}}\,\hat{\mathbf{\mathbf{e}}}_{x} + t_{yy} \sin{\theta_{i}^{p}} \,\hat{\mathbf{\mathbf{e}}}_{y} \right \}.
\end{equation}
Plugging the preceding amplitude-vectors into Eq. (\ref{TIME_AVERAGED_POYNTING_VECTOR_OF_THE_REFLECTED_EM_WAVE}) and Eq. (\ref{TIME_AVERAGED_POYNTING_VECTOR_OF_THE_TRANSMITTED_EM_WAVE}), followed by feeding the output into the definitions of reflectance $R$ and transmittance $T$ yields:
\begin{equation}\label{TRANSMITTANCE}
T  \! \equiv \! \frac{\left| \langle\mathbf{S}_{t}\rangle \right|}{\left| \langle\mathbf{S}_{i}\rangle \right|} \! =\! \left[ \! \frac{\varepsilon^{r}_{2} \, \mu^{r}_{1}}{\varepsilon^{r}_{1} \, \mu^{r}_{2}} \! \right]^{\!\!\frac{1}{2}} \!\! \left\{ |t_{xx}|^{2} \cos^{2}{\!\theta_{i}^{p}} + |t_{yy}|^{2} \sin^{2}{\!\theta_{i}^{p}} \right\}
\end{equation}
\begin{equation}\label{REFLECTANCE}
R \! \equiv \! \frac{\left| \langle\mathbf{S}_{r}\rangle \right|}{\left|  \langle\mathbf{S}_{i}\rangle \right|} \! =\! |r_{xx}|^{2} \cos^{2}{\!\theta_{i}^{p}} + |r_{yy}|^{2} \sin^{2}{\!\theta_{i}^{p}}.
\end{equation}
The fraction of the incident EM flux dissipated into the electrically-conductive interface, is then referred to as the absorbance and quantified by $A \! \equiv \! 1-(R\!+\!T)$. Plugging the Fresnel reflection (transmission) coefficients which are (implicitly) given by Eq. (\ref{FRESNEL_REFLECTION_COEFFICIENTS_NORMAL_INCIDENCE_XX_AND_YY}) into the definition of optical absorbance $A$ yields the expression given by Eq. (\ref{OPTICAL_ABSORPTION_OF_ANISOTROPIC_2D_ELECTRON_GAS}), if only the terms that are proportional to $\alpha$ are retained. Also, the substitution of $\phi \equiv \theta_{\mathbf{i}}^{p} - \theta_{d}$ for $\theta_{i}^{p}$ recovers the formalism for a general direction of the drift velocity given by $\hat{\mathbf{v}}_{\!d}= \hat{\mathbf{\mathbf{e}}}_{x} \cos{\theta_{d}} + \hat{\mathbf{\mathbf{e}}}_{y} \sin{\theta_{d}}$.
\renewcommand{\theequation}{E\arabic{equation}}
\setcounter{equation}{0}
\section{Derivation of the Kerr and Faraday rotation angles}\label{Appendix:E}
The electric field corresponding to the \textit{w}-wave (\textit{w}= \textit{r},\textit{t}) at $z \! = \! 0$ plane (where the graphene sheet is located) can be formally expressed as follows:
\begin{equation}\label{JONES_FORM_OF_THE_W_WAVE}
\mathbf{\mathbf{E}}_{w} (z \! =\! 0,t)=\left|\overline{\mathbf{\mathbf{E}}}_{w}\right| \Re{ \left[ e^{-i\omega t} \!\! \sum_{n=x,y} {\!\! e^{i\alpha_{w}^{n}} \cos{\xi_{w}^{n}} \, \hat{\mathbf{\mathbf{e}}}_{n}}\right] },
\end{equation}
where $\alpha_{w}^{n}\!\equiv\!\arg{\!\left(\overline{\mathbf{\mathbf{E}}}_{w} \! \cdot \! \hat{\mathbf{\mathbf{e}}}_{n}\right)}$ and $\cos{\xi_{w}^{n}}\! \equiv \! \left|\overline{\mathbf{\mathbf{E}}}_{w} \! \cdot \! \hat{\mathbf{\mathbf{e}}}_{n}\right|/\left|\overline{\mathbf{\mathbf{E}}}_{w}\right|$ ($\cos{\xi_{w}^{y}} = \sin{\xi_{w}^{x}}$). The magnitude of the electric-field vector of the \textit{w}-wave at $z \! = \! 0$ is then given by
\begin{figure}
	\begin{center}
		\includegraphics[clip,width=0.98\columnwidth]{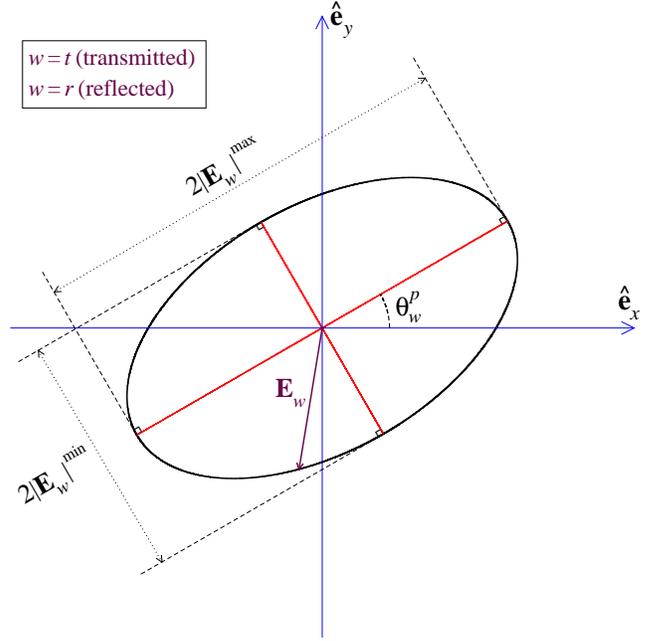}
		\caption{(Color online) Illustration of the ellipse being traced out by the tip of the electric-field vector of the \textit{t} or \textit{r} waves at an arbitrary $z=z_{\textsc{\tiny{0}}}$ plane. For a general $\theta_{i}^{p}$, the ellipses corresponding to each of the \textit{t} and \textit{r} waves are \textbf{not} identical.}
		\label{ELLIPTIC_POLARIZATION_OF_THE_REFLECTED_AND_TRANSMITTED_EM_WAVES}
	\end{center}
\end{figure}
\begin{equation}\label{MAGNITUDE_OF_THE_W_WAVE_ELECTRIC_FIELD}
\left|\mathbf{\mathbf{E}}(z \! =\! 0,t)\right|\! = \! \left|\overline{\mathbf{\mathbf{E}}}_{w}\right| \! \left[\sum_{n=x,y}{\!\! \left[\cos{\!\left(\omega t \! - \! \alpha_{w}^{n}\right)} \cos{\xi_{w}^{n}} \right]^{2}} \right]^{\!\!\frac{1}{2}}\!\!.
\end{equation}
As it can be seen from the geometrical details presented in Fig. \ref{ELLIPTIC_POLARIZATION_OF_THE_REFLECTED_AND_TRANSMITTED_EM_WAVES}, the polarization angle of the \textit{w}-wave is the angle between the $x$-axis and the electric-field vector of the \textit{w}-wave when the magnitude is maximal. Setting the time-derivative of $\left|\mathbf{\mathbf{E}}(z \! =\! 0,t)\right|$ equal to zero:
\begin{equation}\label{DERIVATIVE_OF_MAGNITUDE_OF_THE_W_WAVE_ELECTRIC_FIELD}
\left[\!\frac{d\left|\mathbf{\mathbf{E}}(z \! =\! 0,t)\right|}{dt} \right]_{t=t_{\textsc{\tiny{0}}}} \!\!\!\!\!\! = \! 0 ,
\end{equation}
yields the condition for the maximal time $t \! = \! t_{\textsc{\tiny{0}}}$, 
\begin{equation}\label{CONDITION_FOR_THE_MAXIMAL_TIME}
\sum_{n=x,y}{\!\!\sin{[2(\omega t_{\textsc{\tiny{0}}} \! - \! \alpha_{w}^{n})]} \cos^{2}{\!\xi_{w}^{n}} }=0,
\end{equation}
which leads us to the following relation:
\begin{equation}\label{RELATION_RESULTED_FROM_SETTING_THE_DERIVATIVE_EQUAL_TO_ZERO}
\tan^{2}{\!\xi_{w}^{x}}=-\frac{\sin{ [\omega t_{\textsc{\tiny{0}}}-\alpha_{w}^{x}]}\cos{ [\omega t_{\textsc{\tiny{0}}}-\alpha_{w}^{x}]}}{\sin{ [\omega t_{\textsc{\tiny{0}}}-\alpha_{w}^{y}]}\cos{ [\omega t_{\textsc{\tiny{0}}}-\alpha_{w}^{y}]}}.
\end{equation}
On the other hand, the polarization angle of the \textit{w}-wave can be expressed using the Jones form given by Eq. (\ref{JONES_FORM_OF_THE_W_WAVE}):
\begin{equation}\label{TANGENT_OF_THE_POLARIZATION_ANGLE_OF_THE_W_WAVE}
\tan{\theta^{p}_{w}}=\frac{\cos{ [\omega t_{\textsc{\tiny{0}}}-\alpha_{w}^{y}]} \, \cos{\xi_{w}^{y}}}{\cos{ [\omega t_{\textsc{\tiny{0}}}-\alpha_{w}^{x}]} \, \cos{\xi_{w}^{x}}},
\end{equation}
which can be recast into the following form
 \begin{equation}\label{TANGENT_MIDDLE_FORM}
\tan{[2\theta^{p}_{w}]}=\frac{\left[1-\tan^{2}{\!\xi_{w}^{x}}\right] \tan{[2\xi_{w}^{x}]}}{\frac{\cos{ [\omega t_{\textsc{\tiny{0}}}-\alpha_{w}^{x}]}}{\cos{ [\omega t_{\textsc{\tiny{0}}}-\alpha_{w}^{y}]}}-\tan^{2}{\!\xi_{w}}\frac{\cos{ [\omega t_{\textsc{\tiny{0}}}-\alpha_{w}^{y}]}}{\cos{ [\omega t_{\textsc{\tiny{0}}}-\alpha_{w}^{x}]}}}.
\end{equation}
Plugging the expression for $\tan^{2}{\!\xi_{w}^{x}}$ given by Eq. (\ref{RELATION_RESULTED_FROM_SETTING_THE_DERIVATIVE_EQUAL_TO_ZERO}) into Eq. (\ref{TANGENT_MIDDLE_FORM}) together with the application of a number of simple trigonometric identities yields the final relation:
\begin{equation}\label{TANGENT_FINAL_FORM}
\tan{[2\theta^{p}_{w}]}=\tan{[2\xi_{w}^{x}]} \cos{[\alpha_{w}^{x}-\alpha_{w}^{y}]},
\end{equation}
whose insertion into the following trigonometric identity
\begin{equation}\label{TRIGONOMETRIC_IDENTITY_FOR_TAN(2X)}
\tan{\left[2\left(\theta^{p}_{w}-\theta^{p}_{i}\right)\right]} = \frac{\tan{[2\theta^{p}_{w}]}-\tan{[2\theta^{p}_{\mathbf{i}}]}}{1+\tan{[2\theta^{p}_{\mathbf{i}}]}\tan{[2\theta^{p}_{w}]}} ,
\end{equation}
yields the relation given by Eq. (\ref{KERR_ANGLE}). Also, feeding the expression for the amplitude-vector of the \textit{r}-wave given by Eq. (\ref{THE_AMPLITUDE_OF_THE_REFLECTED_EM_WAVE}) into the definition of $\alpha_{r}^{n}$ and $\xi_{r}^{x}$ yields:
\begin{equation}\label{COSINE_OF_JONES_ANGLE_REFLECTED_WAVE_EXPLICIT}
\tan{\!\xi_{r}^{x}} \equiv \frac{\left|\overline{\mathbf{\mathbf{E}}}_{r} \! \cdot \! \hat{\mathbf{\mathbf{e}}}_{y}\right|}{\left|\overline{\mathbf{\mathbf{E}}}_{r} \! \cdot \! \hat{\mathbf{\mathbf{e}}}_{x}\right|} = \frac{|r_{yy}|}{|r_{xx}|} \tan{\theta_{i}^{p}}
\end{equation}
\begin{equation}\label{ARGUMENT_OF_THE_REFLECTED_WAVE}
\psi_{r} \equiv \alpha_{r}^{x}-\alpha_{r}^{y} = \arg{\!\left(\frac{\overline{\mathbf{\mathbf{E}}}_{r} \! \cdot \! \hat{\mathbf{\mathbf{e}}}_{x}}{\overline{\mathbf{\mathbf{E}}}_{r} \! \cdot \! \hat{\mathbf{\mathbf{e}}}_{y}}\right)} = \arg{\!\left(\frac{r_{xx}}{r_{yy}}\right)}.
\end{equation}
The expressions for $\tan{\xi_{t}^{x}}$ and $\psi_{t}$ can be obtained simply through substituting the Fresnel transmission coefficients for the reflection coefficients in Eqs. (\ref{COSINE_OF_JONES_ANGLE_REFLECTED_WAVE_EXPLICIT}) and (\ref{ARGUMENT_OF_THE_REFLECTED_WAVE}).
\clearpage
\bibliography{Draft_BiB_TS}
\end{document}